\input harvmac
\input epsf

\def\nup#1({Nucl.\ Phys.\ $\bf {B#1}$\ (}
\noblackbox
\Title{\vbox{
\hbox{HUTP-97/A103}
\hbox{UCSB-97-23}
\hbox{\tt hep-th/9712048}
}}{Branes and Fundamental Groups}
\bigskip
\centerline{Rajesh Gopakumar $^{(1,2)}$ and  Cumrun Vafa$^{(2)}$}
\bigskip
\centerline{$^{(1)}$ Dept. Of Physics, University of California}
\centerline{Santa Barbara, CA 93106}
\smallskip
\smallskip
\centerline{$^{(2)}$ Lyman Laboratory of Physics}
\centerline{Harvard University}
\centerline{Cambridge, MA 02138}

\vskip .3in
We consider bound states of D-branes wrapped around cycles with non-trivial
fundamental groups of finite order.  We find a new mechanism for binding D-branes
by turning on flat discrete abelian and non-abelian gauge fields on their
worldvolume.  As a concrete application we study
type IIB in the background where an $S^3/G$ shrinks, where $G$
is a discrete subgroup of $SU(2)$ acting freely on $S^3$.

\goodbreak

\Date{\vbox{\hbox{\sl {December 1997}}}}

\lref\strom{A. Strominger, ``Massless Black Holes and Conifolds in String
Theory'', \nup451 ,(1995), 96}

\lref\pol{J. Polchinski, ``Dirichlet-Branes and Ramond-Ramond Charges'',
Phys.Rev.Lett. {\bf 75}(1995), 4724}

\lref\witb{E. Witten, ``Bound States Of Strings And $p$-Branes'',
\nup460 (1996), 335}

\lref\bsv{M. Bershadsky, V. Sadov and C. Vafa, ``D-Branes and Topological
Field Theories'', \nup463 (1996), 420}

\lref\witef{E. Witten, ``Dyons of charge ${e\theta \over 2\pi}$'',
Phys.Lett. {\bf B86}(1979), 283}

\lref\mgst{B. Greene, D. Morrison and A. Strominger, ``Black Hole Condensation 
and the Unification of String Vacua'', \nup451 (1995), 109}

\lref\gmv{B. Greene, D. Morrison and C. Vafa,
``A Geometric Realization of Confinement'', \nup481 (1996), 513}

\lref\seiin{K. Intriligator, N. Seiberg, ``Mirror Symmetry in
Three Dimensional Gauge Theories'',
Phys.Lett. {\bf B387} (1996), 513}

\lref\hov{K. Hori, H. Ooguri and C. Vafa,
``Non-Abelian Conifold Transitions and N=4 Dualities in Three Dimensions''
\nup504 (1997), 147}

\lref\asmo{P. Aspinwall, D. Morrison, ``Chiral Rings Do Not Suffice:
N=(2,2) Theories with Nonzero Fundamental Group'', Phys. Lett. {\bf B334}
(1994), 79}

\lref\wicom{E. Witten, ``Comments on String Dynamics'', Proceedings
of Strings 95.}

\lref\dtv{C. Vafa, ``Modular Invariance and Discrete Torsion on Orbifolds'',
\nup273 (1986), 592}

\lref\vt{C. Vafa, ``A Stringy Test of the Fate of the Conifold''
\nup447 (1995), 252}

\lref\bcov{M. Bershadsky, S. Cecotti,H. Ooguri, C. Vafa ,
``Kodaira-Spencer Theory of Gravity and Exact Results for
Quantum String Amplitudes'', Commun.Math.Phys. 165 (1994), 311}

\newsec{Introduction}

One of the most beautiful aspects of our recent insight into
non-perturbative aspects of string theory involves how branes
wrapped around vanishing cycles can lead to new low energy degrees
of freedom.  The particular case of type IIB with a vanishing $S^3$
in particular leads to a massless hypermultiplet corresponding to
a 3-brane wrapped around it \strom .  Due to
the identification of D-branes as the sources of RR charge \pol\
one can gain more insight into questions involving bound states
of D-branes \witb .  In particular one can show
\bsv\ that multiple wrappings of 3-branes around
$S^3$ do not lead to any bound states, as anticipated in \strom .

In this note we consider D-branes wrapped around cycles with
non-trivial fundamental group $G$ of finite order.  We will show that
for each non-trivial irreducible
representation of $G$ of dimension $d_i$, we obtain
a new bound state of $d_i$ D-branes wrapped around the same
cycle.  The basic idea is to turn on a non-trivial discrete flat gauge
field in the D-brane worldvolume.  Note that this can be done
due to the fact that for $d_i$ coincident D-branes we obtain
a $U(d_i)$ gauge symmetry. The condition that the representation
be {\it irreducible} gets translated to having {\it bound states}
of  D-branes.   We illustrate the basic results in the context
of type IIB compactifications on Calabi-Yau threefolds where an
$S^3/G$ vanishes, where $G$ is a subgroup of $SU(2)$ freely acting
on $S^3$.  In particular, in this example we find that the
well-known correspondence
between discrete subgroups of $SU(2)$ and affine Dynkin diagrams is
rather useful in giving the spectrum of bound states.  Namely,
for each subgroup of $SU(2)$ the number of bound states is
equal to the number of nodes of the corresponding affine Dynkin
diagram.  Moreover the Dynkin index corresponding to each node
signifies the number of D-branes of which this is a bound state.

\newsec{D-branes and non-trivial fundamental group}

Let us consider type IIA/B strings compactified on some manifold
$M$.
Consider a compact submanifold $C\subset M$.  Let us
assume that it is a supersymmetric cycle.
Moreover let us assume that it has a non-trivial fundamental group:
$$\pi_1 (C)=G.$$
We will assume that $G$ is finite\foot{The considerations of this
paper can be easily generalized to the more general case where $G$
has isolated finite dimensional unitary representations.}.  
If we consider a D-brane
(of the appropriate dimension in the appropriate theory)
wrapped around $C$ we obtain a particle in some supersymmetry
multiplet, depending on the theory we are dealing with.  However
we have more degrees of freedom available to us:  On the D-brane
worldvolume we have a $U(1)$ gauge symmetry.  The vacuum
configurations of the D-brane correspond to particles.
However the existence of a non-trivial fundamental group introduces
a new subtlety.  The vacuum equation on the D-brane worldvolume implies
$$F=0.$$
This equation
can be solved in more than one way.  The simplest way
is to set $A=0$, but we can also set $A$ to be a non-trivial
flat $U(1)$ connection.  The number of inequivalent ways this can
be done is equal to the number of inequivalent 1 dimensional
representations of $G$.  To see this, let $p_\gamma$
denote a one dimensional path in $C$ representing each $\gamma \in G$.
Let
$$g_\gamma ={\rm exp}(i \int_{p_\gamma} A)$$
denote the wilson line for that path, where we are assuming $F=dA=0$.
Note that due to flatness, this does not depend on which representative
path we take.  Then $g_\gamma$ satisfy the same relations as the fundamental
group $G$ by composing the holonomies along the composed paths.
  In other words we can view $\{ g_{\gamma} \} $ as forming a one dimensional
representation of $G$.  We thus conclude that the number of inequivalent
D-brane states wrapped once around $C$ is equal to the number of
one dimensional representations of $G$.\foot{Note that if $C$ had a non-trivial
$S^1$, then $G$ would have
${\bf Z}$ as a subgroup.  In this case one {\it does not} get
new bound states.  This is because turning on Wilson lines
around $S^1$ have a continuous parameter associated with it, and
to find the bound states we need to find the ground states of the
quantum mechanics problem on the dual circle.  As is well known (apart
from the usual fermionic degrees of freedom) this is one dimensional.}
Note that, even though the argument about the ground states
we gave is classical, the masses of the corresponding D-branes
are protected quantum mechanically by the BPS condition.  Thus the classical
arguments suffice for the existence of these BPS ground states.

Now let us consider $N$ D-branes wrapped around $C$.  Let us assume
that the dynamics is such that without turning on gauge fields
there are no multi-wrapped
bound states (as is the case for example for the conifold \bsv ).
Now let us ask if there are any new bound states due to the existence of
a non-trivial fundamental group on $C$.  In this case we have a $U(N)$
gauge theory on the D-brane worldvolume.  In solving the flatness condition
$F=0$ we can, as in the case with a single D-brane, turn on non-trivial
flat connections.  For the same reason as before the number of inequivalent
ways this can be done corresponds to the number of inequivalent $N$-dimensional
representations of $G$.  However, the reducible representations
will not correspond to a bound state:  The fact that we have reducible
representation means that we can find at least one
$U(1)$ subgroup of $U(N)$ which commutes with the turned on
wilson lines.  This in particular means that the scalars $\phi$ in that
$U(1)$ multiplet are massless.  This is because the
D-brane action
which couples the fields to the background gauge field $A_c$
does so by group commutators
\eqn\comu{S=...+\int \big| [\phi ,A_c]\big|^2}
and this is zero for the choice of $U(1)$ direction we have made.  The
massless scalar mode $\phi$ is responsible for seperating D-branes
(dividing D-branes in some particular way).
  Thus the same
arguments that shows that multi-wrapped states are not bound, applied to this
$U(1)$, show that the reducible representations do not give rise
to bound states.  Let us instead assume that $N=d_i$ where $d_i$
is the dimension of some {\it irreducible} representation of $G$.
In this case we can identify the holonomies of the $U(N)$ gauge
theory with this irreducible representation of $G$.  The very fact
that the representation is irreducible, means that no massless
deformation
exists, i.e. all commutators of the form \comu\ are non-trivial
and give mass to all such directions, except for the overall $U(1)$ which
corresponds to the center of mass deformation.  We thus automatically
get a bound state.  We conclude that the number of bound
states are in one to one correspondence with the number
of irreducible representations
of $G$ which in turn is equal to the number
of conjugacy classes of $G$.  Moreover the dimension of the representation is
the number
of D-branes wrapped around $C$.

\subsec{Discrete quantum numbers and D-brane/string bound states}

One way one obtains a discrete non-trivial fundamental group is
by modding a simply connected space by some free group action $G$.
In this case the fundamental group of the space is $G$.
If we are dealing with type II string theories compactified on such
spaces, we get the familiar twisted sectors of strings, which
correspond to strings which are closed up to the $G$ action.
We get twisted sectors, one for each conjugacy
class of $G$.  Moreover the interaction between them is consistent
with the group multiplication for all the elements in a
conjugacy class.   We can define a discrete symmetry
which is consistent with this interaction (in orbifold
constructions this is known as
a `quantum symmetry'):  Let $\hat G$ be the abelianization
of $G$ (i.e. ${\hat G}=G/[G,G]$).  Then for each conjugacy class $[g]\in G$ we
get an element
of $\hat G$ (this is not a one to one map in the general case).  Then the
string interactions are consistent with the abelian multiplication of
$\hat G$, and we can thus associate a $\hat G$ symmetry to strings.
In the case $G$ is abelian, $\hat G=G$ and the symmetries of the
orbifold theory we have
are in fact the same as $G$.  For example if $G={\bf Z}_n$,
in the orbifold theory we also get a new ${\bf Z}_n$ symmetry,
which is simply the conservation law associated
with the labels of the twisted sectors.  This is reflected
by associating a phase ${\rm exp}(2\pi i k/n)$ to states coming from
the $k$-th twisted sector.
  In fact we can summarize the
general case by saying that the
discrete symmetries of the orbifold theory
correspond to choosing a representation of $\hat G$ which
is the same as finding 1-dimensional representations of $G$.

It is natural to ask whether the wrapped D-branes we have discussed
above carry charge under this $\hat G$.  Let us first do this
for the case of $G={\bf Z}_n$.  It is convenient to consider the case
where $n\rightarrow \infty$, i.e. the case where the
group is ${\bf Z}$ and view the other cases as
embeddable in it.  In particular
let us consider $S^1=R/{\bf Z}$.  In this case the corresponding
notion of the twisted sectors are just the winding sectors,
and the new $Z$ charge is just the winding number charge.
Let us denote the winding direction by 1.   A winding string is charged
under the $B_{01}$ field.  Consider a D-brane which is wrapped
around this circle.  Then considering a winding number $n$
state of string, amounts to considering a state with electrical
flux of $n$ units \witb , i.e, $dA_1/dt=n$.
Note that this is the conjugate variable to the wilson line $A_1$.
Let us denote the Wilson lines by $A_1=\theta$.  Then we can view a sector
with a given Wilson line as a superposition of the electrical flux states
(in the usual position/momentum representation):
$$|\theta \rangle =\sum {\rm exp} (in \theta) |n\rangle$$
Indeed, in terms of a T-dual (in direction 1) picture, $\theta$ is the center
of mass position and the winding $n$ goes over to the momentum around
the dual circle.
Thus the conservation law associated with the addition in the
winding number, gets mapped to the shift symmetry in the $\theta$ angle
which is the Wilson line on the D-brane world volume.  In other words,
the symmetry
$$|n\rangle \rightarrow {\rm exp}(in \alpha)|n\rangle$$
gets mapped to
$$|\theta \rangle  \rightarrow |\theta +\alpha \rangle$$
The ${\bf Z}_n$
version, which is the case of interest for us, is now clear:  If we consider  
the symmetry
which corresponds to multiplying the first ${\bf Z}_n$ twisted
state of string by ${\rm exp}(2\pi i/n)$ it acts
on the D-brane world volume by turning on one unit
of ${\bf Z}_n$ Wilson line.  In other words
it permutes the D-brane bound states.
In order to get eigenstates of this action of the quantum ${\bf Z}_n$
charge operator, we can
take a superposition of D-branes with discrete Wilson lines
turned on.   A natural question arises as to whether
we can describe these superpositions in a classical way.  In
fact from what we said before this can be partially done, namely
we can turn on discrete electrical fluxes which would correspond
to these eigenstates.  In fact we can also
represent the electric fluxes as the twisted string states themselves.
Thus the D-brane states with a superposition of different discrete Wilson
lines can be viewed as bound states of D-branes without
any Wilson line, with twisted strings.
The actual ground state looks very different
from a string attached to the D-brane--in particular it 
has the same mass, by the BPS condition, as the D-brane without the
string attached.  The ground state is a twisted
string completely `dissolved' in the D-brane.

 This can be easily generalized
to include the non-abelian case:
Each D-brane bound state is characterized by an irreducible
representation of $G$.  The action of a particular
symmetry of $\hat G$, corresponding to a one dimensional
representation $R_1$ of $G$, on the wrapped
D-brane state corresponding to representation $R$
is given by
$$R\rightarrow R\otimes R_1$$
(This is a shift in the $U(1)$ part of the $U(N)$ holonomy -- precisely 
the action of the generator of the $\hat G$ symmetry.)
This in general permutes the corresponding D-brane bound states given
by specific Wilson lines. Clearly we can choose a linear superposition
of D-brane bound states to form eigenstates of this action,
and again these will correspond to bound states of
D-branes associated with the representation $R$ with a twisted
string state, as in the abelian case.

\newsec{Examples based on Calabi-Yau threefold}
As a concrete application of the above ideas let us consider the
case of type IIB compactification on Calabi-Yau threefolds, and let
us consider the limit where the threefold develops a conifold singularity.
This means that locally the manifold looks like
$$z_1z_4-z_2z_3= \mu .$$
As $\mu \rightarrow 0$ the manifold is singular at $z_1=z_2=z_3=z_4=0$.
In this example one has a three sphere which vanishes in this limit.
One way to see the $S^3$  is to consider (assuming $\mu$ is real)
the locus where $z_4=z_1^{*}$ and $z_3=-z_2^{*}$.  To study this
space in more detail we
consider the $2\times 2$ matrix
\eqn\amat{\eqalign{A=\pmatrix{z_1 & z_3 \cr
z_2 & z_4 \cr}}}
In terms of this, the local model for the conifold is given by
$${\rm det}(A)=\mu$$
There is an $SL(2,{\bf C})_L\times SL(2,{\bf C})_R$
acting on the conifold space by left and right multiplication
of $A$:
$$A\rightarrow K_L A K_R$$

If we consider the 3-brane wrapped around $S^3$, then
we get a hypermultiplet charged under $U(1)$,
with mass proportional to
$\mu$
\strom , where from the gauge theory viewpoint $\mu$ is
the vev of the scalar in the $N=2$ $U(1)$ vector multiplet.
  As $\mu \rightarrow 0$ we obtain a massless hypermultiplet.
One signature of the massless hypermultiplet is the fact that the
coupling constant of the $U(1)$ near $\mu \rightarrow 0$ is
given by $\tau={1\over 2
\pi i}
{\rm log}\mu$ which is what one expects from the 1-loop correction
for $N=2$ QED with a hypermultiplet of mass $\mu$.  Note that the  coupling
constant
undergoes a monodromy $\tau \rightarrow \tau +1$
as $\mu \rightarrow \mu \!\ {\rm exp}(2\pi i)$.  In other words
the $\theta$ angle shifts by $2 \pi$.
This in particular
means that the magnetic states of charge one become dyonic
as we go around the origin in $\mu$ space.  This is in fact the
Witten effect \witef.  Geometrically this is realized rather
nicely by the fact that a magnetic state of charge one is represented
by a 3-cycle which intersects this $S^3$ at one point. Let us call this
class $[b]$.  Moreover let ue denote the class of the $S^3$ cycle
by $[a]$. Then as $\mu$ goes around the origin we come back to the
same geometry but the cycles undergo the monodromy
$$[b]\rightarrow [b]+[a]$$
Thus the magnetic state picks up one unit of electric charge.

We now consider modding out this space by a discrete subgroup
of $SU(2)$.  There are a number of ways this can be done depending
on how one embeds $SU(2)$ in the $SL(2,{\bf C})_L\times SL(2,{\bf C})_R$.
For the purposes of this paper, it suffices to consider the $SU(2)$
identified with the $SU(2)$ subgroup of $SL(2,{\bf C})_L$.
Let $G$ denote a discrete subgroup of $SU(2)$.  Then modding out
the conifold by $G$:
$$A\rightarrow G\cdot A$$
gives rise to a new space, which again satisfies the Calabi-Yau
condition.  Note that this action is free, i.e., we have no
fixed points.  This in particular means that the new space will
have a fundamental group isomorphic to $G$.  Note also that the
$S^3$ is mapped to itself under the $G$ action.  Now consider the
3-cycle
$$C=S^3/G$$
Note that $\pi_1(C)=G$.  Now the situation is similar to what we had
before modding out, except that the minimal three cycle charge is now
given by a 3-brane wrapped $1/|G|$ times around the $S^3$,
where $|G|$ denotes the order of $G$. In particular
the basic unit of charge is reduced by a factor of $1/|G|$.
Now we will apply the considerations of the previous section and deduce
the bound state spectrum of the D3 branes wrapped around $C$.

In order to do this, it is convenient to first review
some basic facts about the $SU(2)$ subgroups.  They are in one to
one correspondence with A-D-E.  The cyclic subgroup of $SU(2)$
is identified with the A series, the Dihedral subgroup with D and
the exceptional subgroups with the E series (this is familiar
from how the gauge symmetries arise in modding out ${\bf C}^2$ by
the corresponding discrete subgroup of $SU(2)$).  This correspondence
has an interesting further aspect:  If we consider the {\it affine}
Dynkin diagram associated with each subgroup $G$ of $SU(2)$, the number
of nodes (which is equal to the rank of the corresponding Lie group plus
one) is equal to the number of irreducible representations of $G$. Moreover
the dimension of the corresponding representation is in 1-1 correspondence
with the Dynkin index associated with that node (see figure 1).

%

\bigskip
\centerline{\epsfxsize 2.truein \epsfysize 2.truein\epsfbox{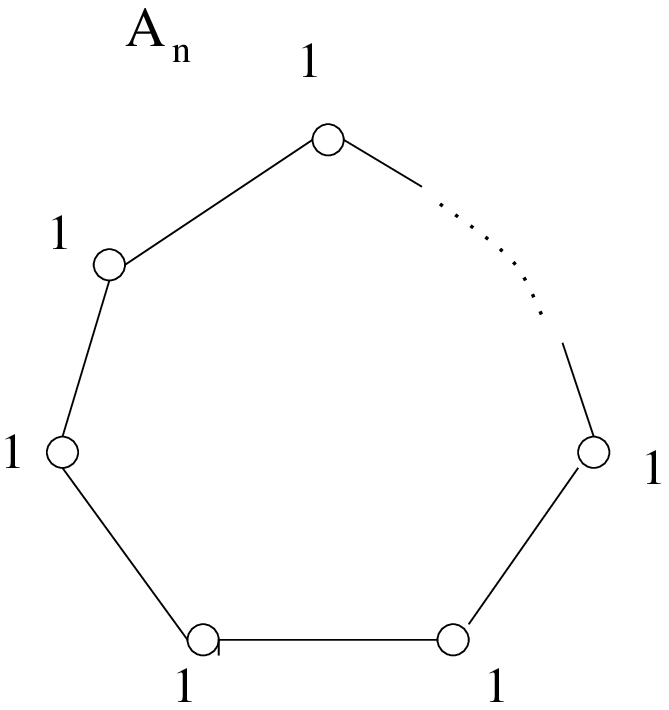}}
\smallskip
\centerline{\epsfxsize 2.truein \epsfysize 1.truein\epsfbox{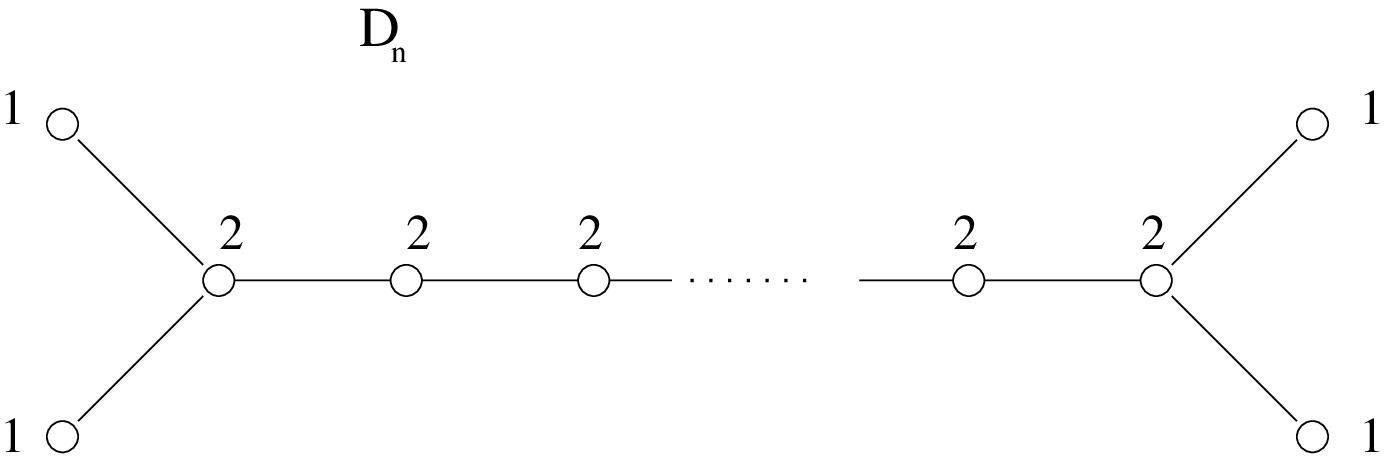}}
\smallskip
\smallskip
\centerline{\epsfxsize 2.5truein \epsfysize 2.5truein\epsfbox{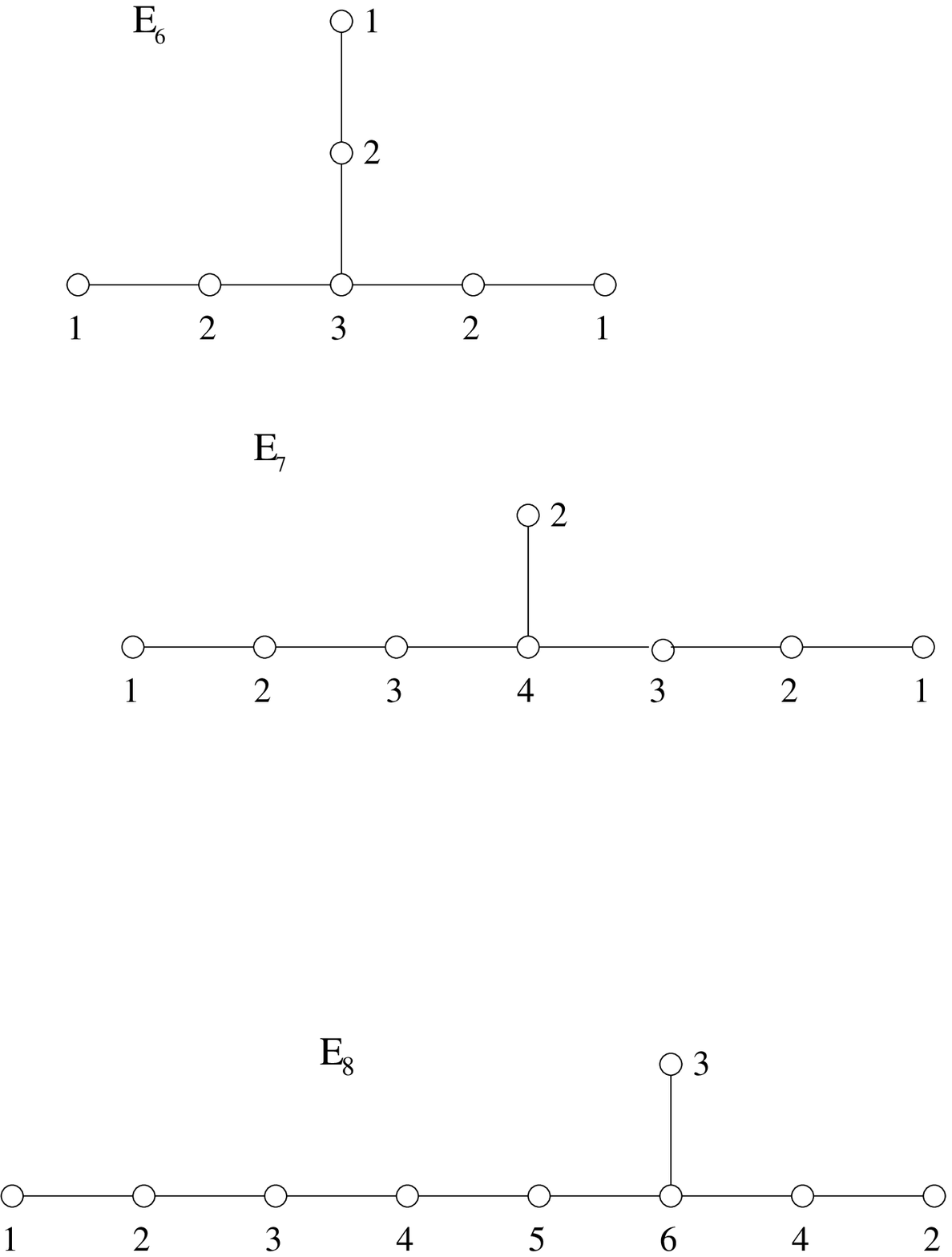}}
\leftskip 2pc
\rightskip 2pc
\noindent{\ninepoint\sl \baselineskip=8pt {\bf Fig.1}: {\rm
The A-D-E Affine Dynkin diagrams encode the information
about the representations of cyclic, dihedral and exceptional
subgroups of $SU(2)$.  Each node is associated
with an irreducible representation, and the Dynkin index
corresponds to the dimension of the representation.}}

In fact one can also deduce the tensor product of the fundamental
representation of $SU(2)$ with any of the other representations
from Dynkin diagrams:
Tensoring the fundamental representation with an irrep corresponding to a  
node yields the direct
sum of the irreps associated with all the adjacent nodes.

Coming back to our problem of finding the number of D3-brane states
wrapped around $C$, using the result of the previous section
we see that they are in one to one correspondence
with the irreps of $G$ which are in turn in one to one correspondence
with the nodes of the corresponding affine A-D-E, with the Dynkin
index denoting the number of D3 branes wrapped around $C$.
Note that this number is also the same as the $U(1)$ charge
of the modded theory (the minimal charge is $1/|G|$ times
the minimal $U(1)$ charge of the unmodded theory).  Also note
that the BPS condition implies that their mass is given by
$M_i=\mu d_i/|G|$, where $d_i$ denote the Dynkin indices.

\subsec{Various Checks of the Results}
We now pass the previous results through some checks.
First let us ask what should the correction be to the effective
coupling of the $U(1)$ theory, normalized so that the basic
electric unit is $1/|G|$ times the unmodded theory.  To see
that it suffices to see how a magnetic state will undergo monodromy
as $\mu$ goes around the origin.   The magnetic state, given by the
cycle $[b]$ in the unmodded theory, continues to be the minimal
unit in the modded theory as well (because it intersect the $S^3/G$
at one point).  Thus the same monodromy continues to hold, namely
$[b]\rightarrow [b]+[a]$ but now $[a]$, which is the $S^3$ class,
does not represent the minimal charge state for the new $U(1)$ but
$|G|$ times the minimal unit.  Thus the $\theta$ angle for the new $U(1)$
undergoes a monodromy $2\pi |G|$.  This should be reproducible
as a result of the contribution of the light modes.  In the above
we have argued that there are as many light modes as the nodes
of the corresponding affine Dynkin diagram and that each will carry
a charge $q_i=d_i$ which is the same as the Dynkin index associated
with the corresponding node, with mass $M_i =\mu d_i/|G|$.
The contribution they make to the coupling constant of the $U(1)$ theory
is given by the usual one loop result,
$${\tilde \tau}=\sum_i q_i^2 \!\ {\rm log M_i}=(\sum q_i^2)\!\ {\rm log}\mu +{\rm
const.}$$
Thus we pick up the monodromy, as $\mu$ goes around the origin, given by
$$\sum q_i^2=\sum d_i^2$$
We expected this to be equal to $|G|$, and this is indeed the case:
$$\sum d_i^2 =|G|$$
and expresses the well known relation between the dimension $d_i$
of irreducible representation of a group $G$ with its order $|G|$.
We find this a strong check on our basic conclusion.

Let us consider another check.  One of the basic aspects of the
existence of massless modes corresponding
to wrapped D3 branes around vanishing 3-cycles is that it provides a way
to go to new Calabi-Yau compactifications by giving vevs to them \mgst .
  If we consider the case where an $S^3$ shrinks to zero
size, locally we can consider growing an $S^2$ instead.  This is described
as follows.  Consider the point where $\mu =0$.  In such a case we can
define a complex parameter $\lambda =\lambda_1/\lambda_2$ through the equation
\eqn\blup{\eqalign{\pmatrix{z_1 & z_3 \cr
z_2 & z_4 \cr}\pmatrix{\lambda_1 \cr \lambda_2 \cr}=0}}
Note in particular that the point $z_1=z_2=z_3=z_4=0$ which was singular
is now replaced by the sphere parameterized by $\lambda$, i.e.,
we have blown it up to an $S^2$.  We can
use $z_1,z_2, \lambda$ as parametrizing the blown up space (note
that $z_3=-\lambda z_1 \quad z_4=-\lambda z_2$).
  The size of $S^2$ should be described
by varying the vev of a hypermultiplet, and that is precisely identified
with the massless hypermultiplet coming from the wrapped D3 brane
around the vanishing $S^3$.
In the compact set up
 there are some global obstructions as to the sizes of the $S^2$'s
coming from D-term constraints in field theory (see \gmv\
 for the relation of these constraints to a geometric realization
of confinement); however even in such cases
the existence in the field theory of a degree of freedom which describes
the geometric size of blown up $S^2$ is a check of the fact that
a D3 brane wrapped around a vanishing $S^3$
corresponds to a massless hypermultiplet.

Now let us run this check in the case we are considering\foot{
We would like to thank Eric Zaslow for providing us with this
blow up construction.}:  If we mod
out the conifold by $G$, this acts also on the blown up space. It
is easy to see that the action of $G$ on $\lambda$ is trivial,
and that as before $(z_1,z_2)$ form a doublet under the $G$ action.
So for the blown up space given by $(z_1,z_2,\lambda)$ the ADE subgroups
of $SU(2)$ just act on the two coordinates $(z_1,z_2)$.  Thus the
geometry we are getting is an ADE type singularity, parameterized
by a sphere $\lambda$.  The number of blow up modes in this case
is familiar from the study of the corresponding ALE singularities
and is given by the rank of the corresponding ADE, together with
the mode corresponding to the size of the $S^2$ parametrized by
$\lambda$.  Thus we expect that the number of massless hypermultiplets
be the rank of the corresponding ADE plus one.
 This is exactly
what we found by the D-brane analysis.\foot{Once we compactify
on an extra circle, and go to the dual circle we end up with
the type IIA description on the same manifold which leads to $N=4$ ADE
gauge symmetries in 3 dimensions.  In fact
this may have interesting
application for 3d $N=4$ mirror symmetries \seiin
along the lines suggested in \hov .}  Note that this is the same
as compactification of type IIB near an ADE singularity over an
$S^2$, down to four dimensions.  In this case one gets
tensionless strings when the spheres shrink \wicom.
However sometimes tensionless strings can look like ordinary
string associated to higgsing of $U(1)$ factors \gmv\ and this is
indeed the case here.

Finally we can pass this through a string one loop test corresponding
to the coefficient of $R^2$ term generated \vt .
The coefficient of this term should be given by
$$F_1 = \sum_i- {1\over 12}{\rm log}\ M_i$$
for hypermultiplets of masses $M_i$.  This can be computed
using the one loop topological string amplitudes \bcov .
There are only two cases where this has been computed\foot{
We are grateful to Albrecht Klemm for pointing these out to us.}
 with a singularity
of the type we are considering.  One corresponds to the case when one has
a vanishing $S^3$ with a free $Z_5$ action \asmo, and in this case
we expect to get a coefficient of $-{5\over 12}{\rm log} [\mu/5]$ and this is
indeed
the case.  And the other \ref\kl{A. Klemm, unpublished.}\
(coming from the $Z_3$ quotient of a bicubic in ${\bf P}^5$) has
an $S^3/Z_3$ shrinking and this gives the expected $-{3\over 12}{\rm log}
[\mu/3]$
contribution to $R^2$ term.

\subsec{Non-trivial fundamental group and massless wrapped $(p,q)$ strings?}

Apart from the 3-branes that are wrapped around the shrinking 3-cycle in
the case we have discussed so far, there are extra states that seem to be
becoming light. These are $(p,q)$ strings of the IIB theory wrapped on any
of the non-contractible loops of $S^3/G$.   We cannot apriori
argue that they are all stable.  However we can use the discrete charges
available to argue some may be stable.  In particular for the fundamental
strings, if we choose them to be the ground states corresponding
to particular discrete symmetry ${\hat G}$, we can show to all order
in string perturbation theory that
they are stable.  However, the
wrapped $(p,q)$ string states, for weak string coupling,
are very massive and
no quantum charge protects their decay to fundamental string states.
The only quantum charge they could carry is that of the discrete
symmetry we have already discussed, and the light fundamental string states
already carry them.

So let us concentrate on the wrapped fundamental string states and consider
the limit where the $S^3/G$ shrinks.  We can ask if they do give rise to
additional massless stable states.
First we should note that our intuition about their
becoming massless is purely at a classical level. Quantum mechanically we
must see if a BPS mass formula protects their mass. However in this case,
the strings are not charged under the Ramond-Ramond gauge field that
couples to the 3-brane. The only
charge they carry is under a discrete gauge group corresponding to the
abelianisation of the fundamental group, as noted before, and this
does not give rise to any BPS mass formula.  At any rate classically one would
naively expect that one ends up with stable massless states as the  
$S^3/G$ shrinks.
We do not believe this happens, for the following reason:
   Even if
we assume the naive classical mass for them is valid,
the physics of the light modes
is dominated by the light wrapped 3-branes whose mass is
proportional to the volume of the 3-cycle. The light strings have classical
masses which are proportional to $1/3$ power of the volume and thus
are much heavier in the limit the volume shrink.  Moreover even though
the ground states of the fundamental string states in the twisted
sectors are protected from decay to all orders in string perturbation
theory by the discrete charge they carry, this is not the case
non-perturbatively, because the wrapped D3 branes not only are lighter
but also carry the same discrete charges, as discussed before.  Thus a  
fundamental string
can decay to a pair of D3 and anti-D3 branes which are wrapped
(with suitable Wilson lines turned on).
This seems to provide an interesting example for perturbative
string states which are stable to all order in string perturbation,
but nevertheless become unstable in the regime where
string perturbation is not valid (where the $S^3/G$ shrinks).

\newsec{5-Branes}
Up to now we have talked about D-branes wrapped around cycles
with non-trivial $\pi_1$.  Can we do a similar thing for
NS 5-branes of type IIA or M5 brane of M-theory, and
NS 5-brane of type IIB?

For type IIB NS 5-brane the story is the same
as what we have said for D-branes, because there is a gauge
field $A$ living on the worldvolume (after all it is S-dual
to D5 branes).

The story is more subtle for NS 5-branes (or M5 branes) of
type IIA (M-theory).  For a single 5-brane, given that
we have the anti-symmetric tensor field in its world volume,
the question of choosing non-trivial discrete vev's is
exactly the same as the choice of a discrete torsion \dtv.  In particular if  
the spatial part is wrapped around
the 5-cycle $C$, the inequivalent choices for
the vev of $B_{ij}$ correspond to the choice of $H^2(C,U(1))$.
The more subtle question is whether there is anything new that
happens when we have more than one 5-brane.  In this case
there is no classical analog for a non-abelian object such
as $\pi_1$ (note for example that $\pi_2$ is always abelian).
However we will argue, at least for certain cases, that
quantum mechanically there are
additional choices available.  Consider in particular the 5 dimensional
submanifold to be $C=S^1\times {\tilde C}$ and
suppose that $\pi_1({\tilde C})$ is a finite non-abelian group.
In this case for each
$n$ dimensional representation of $\pi_1({\tilde C})$,
 if we consider $n$ 5-branes
wrapped around $C$ we have new bound states corresponding to turning
on certain fields.  To see this we note that $n$ 5-branes wrapped
around $S^1$ give rise to $U(n)$ gauge theory in $4+1$ dimensions.
Thus on ${\tilde C}$ we can turn on non-abelian wilson line,
as we did for the D-brane case in this paper.  This example
shows that even for 5-branes of type IIA (or M-theory) there
are in general non-abelian version of wilson line.  It would
be interesting to see what is the more general version
of this non-abelian vev, when we do not assume that $C$ has
an $S^1$ factor.

We would like to thank A. Klemm, J. Maldacena, A.Strominger 
and E. Zaslow for valuable
discussions.

The research of R.G. was supported by DOE grant 91-ER40618.
The research of C.V. was supported in part by NSF grant PHY-92-18167.

\listrefs

\end